\documentstyle{l-aa}

\newcommand{\be}{\begin{equation}}
\newcommand{\ee}{\end{equation}}
\newcommand{\Hb}{H$\beta$}
\newcommand{\Ha}{H$\alpha$}
\newcommand{\Hd}{H$\delta$}
\newcommand{\Hr}{H$\gamma$}
\newcommand{\HE}{H$\epsilon$}

\begin{document}

\thesaurus{03                   
		(11.19.3;       
		11.19.2;       
		19.37.1)}      

\title {F06296$+$5743 : A Very Massive Starburster ?}

 \author{J.H.Huang
	 \inst{1,3}
 \and Q.S. Gu
	 \inst{1,3}
 \and H.J. Su
	 \inst{2,3}
 \and Z.H. Shang
	\inst{2,3}
	}

 \offprints{J.H.Huang}

  \institute{Astronomy Department, Nanking University, Nanking, China
   \and Purple Mountain Observatory, Nanking, China
  \and United Lab for Optical Astronomy, The Chinese Academy of Sciences, China}

 \date{Received;accepted}
  \maketitle

\begin{abstract}
We have observed an object having strong Balmer absorption lines and blue 
continuum. The inference about the nuclear stellar population drawn from the
line ratios is basically coincident with that drawn from the continuum color, 
both indicating the dominant fraction of early A type population in the nucleus.
This might be expected only if there has been a very massive starburst, burst 
strength of $>$ 10\% at least.

 \end{abstract}

 \section{Introduction}
During our surveying IR-selected Seyferts in 1994-1995, we have observed an 
object having strong Balmer absorption lines, F06296$+$5743.
It is interesting to note that its spectrum exhibits blue continuum.

Far-infrared luminous galaxies with strong Balmer absorption lines have been 
observed before ( see, e.g. Armus, Heckman, \& Miley 1989 ), but their
continuum colors are usually very red. For example, the ratios of the continuum
flux densities at rest wavelengths 6500\AA~and 4800\AA, C65/C48, defined by
Armus, Heckamn, \& Miley ( 1989 ), for their observed three galaxies with Balmer
absorption lines are 1.06, 1.13, and 1.45, respectively, and the equivalent
widths of \Hb~ absorption line are 6.5\AA, 5.2\AA, and 1.2\AA, respectively. As
a comparison, the corresponding values for F06296$+$5743 are 0.77, and 8.2\AA,
respectively.

The anti-correlation between the equivalent widths at \Hb~ and C65/C48 may
imply the effect of the burst strength on the continua,
as pointed out by Bica, Alloin \& Schmidt ( 1990 ). The larger the burst
strength is, the bluer the continuum. Our analyses show that F06296$+$5743
might be a very massive starburster.

\section{Observation and Data Reduction}
 Spectra of F06296$+$5743 were taken on March 4, 1995 at Beijing Astronomical
 Observatory ( BAO ) with 2.16m telescope. The TEK CCD chip 
 ( 1024$\times$350 ) was
 attached to the Carl Zeiss spectrograph. A 300 lines $\rm{mm}^{-1}$ grating
 was used with a 3" wide slit, providing a spectral resolution of 
 10.9\AA~FWHM
 over the range 3800-7600\AA. One pixel corresponds to 4.7\AA. Exposure time
 was 60 min. for object. During our exposure we had a seeing of 1.5". Flux
 calibration was done with spectra of Feige34 and HZ44 ( Barnes \& 
 Hayes 1984 ). FeHeAr-spectra had been taken before and after all the object
 spectra for wavelength calibration.

 Standard reduction procedures in IRAF were followed for bias subtraction,
 flat-field correction, wavelength calibration, extraction, and the flux
 calibration. The terrestrial absorption band of $\rm{O}_{2}$ happened to
 be at the redshifted \Ha~emission line, see Fig 1a. It was removed by the 
 method described in Osterbrock et al ( 1990 ).
 
 The spectrum in Fig 1a exhibits strong Balmer absorption lines. Since this
 absorption is the same at \Ha, \Hb~ and \Hr~ to whithin 30\% ( Kurucz 1979 ),
 our measured equivalent widths at \Hb, \Hr, and \Hd~ also confirm it ( see 
 below ), so we have followed the procedure suggested by Armus, Heckamn \&
 Miley ( 1989 ) to use \Hb~ absorption line to correct  the \Ha~ emission line.

   \begin{figure*}
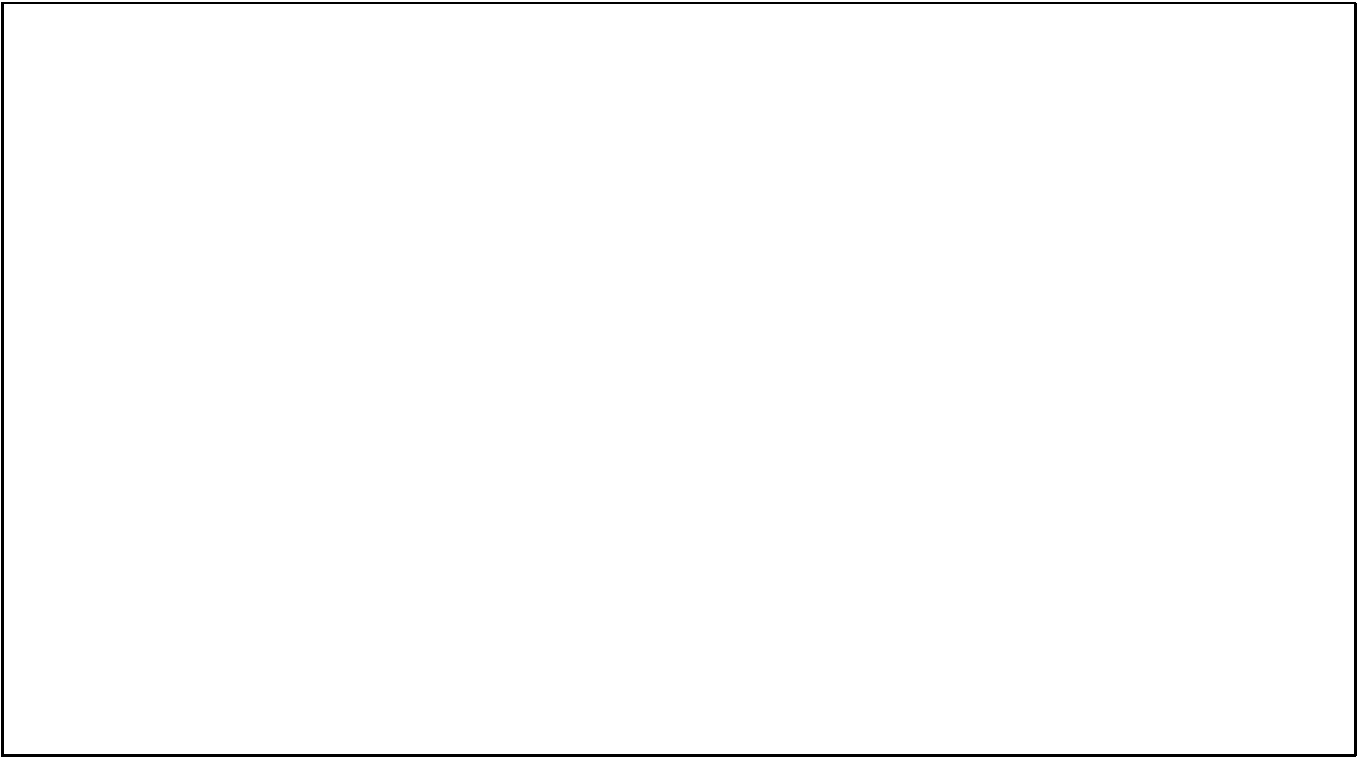

        \picplace{10cm}%
     \caption{Flux-calibrated spectra of F06296$+$5743 showing strong Balmer
     absorption lines, vertical scale is in
    ergs.$\rm cm^{-2} s^{-1} \AA^{-1}$.
   (a) observed flux-calibrated spectrum, no correction was made for
  absorptions and instrumental-broadening; (b) final spectrum, corrected
	 for galactic extinction, absorptions and instrumental-broadening.
	 }
    \end{figure*}

 It is necessary to correct all observed line fluxes for attenuation due to
 reddening, which has two components $-$ foreground reddening due to dust in
 our Galaxy and internal reddening from dust within the source itself. In the
 case of F06296$+$5743, the later one would be difficult to settle, as the
spectrum shows that \Hb~line has no emission component so that it will not be
able to obtain the observed Balmer decrement, which is needed for reducing the 
internal reddening. The foreground reddening could
be accounted for by use of the average Galactic extinction law of Seaton 
( 1979 ) and the extinction estimates towards the direction of F06296$+$5743
given by Burstein \& Heiles ( 1984 ), which is about $\rm A_{B}=$0.29, i.e.
E(B$-$V)$\approx$0.07. The final, dereddened and absorption-corrected spectrum
of F06296$+$5743 is shown in Fig 1b.

\section{Discussion}
\subsection{Nuclear Physical Condition}

The observed emission/absorption line intensities of F06296$+$5743 are tabulated
in Table 1. The redshift derived from \Hb, \Hr~ and \Hd~ lines is 
 $\overline{\rm z}$ = 0.05262. The second column in Table 1 lists the
 wavelengths
 at the rest frame, relating to the wavelengths in the first column. At the 
 distance of  $\overline{\rm z}$ = 0.05262, 3" slitwidth corresponds to an 
 extension of about 3 kpc, with the Hubble constant ${\rm H_{o}}$
 =75 $\rm km/sec/Mpc$.
 So, the equivalent widths and the line intensities
listed in column 4 and 5 are
 those in the nuclear region of 3 kpc in diameter.

 In the case of F06296$+$5743, the line ratios of $\rm [NII]/$\Ha~ and 
 $\rm [OI]6300/$\Ha~ were used instead for classifying the physical condition
 of this source because only appears \Ha~ in emission, while \Hb~ appears in
 absorption. In order to deblend the line of  
 \Ha$\rm +[NII]\lambda\lambda$6548+6583,
 we used the "specfit" in IRAF, an interactive procedure contributed by Gerard
 Kriss, in which the line-width of \Ha~ emission was set to about the same as 
the \Hb~ line. Finally, we found that 
$\rm [NII]/$\Ha = 1.07, and $\rm [OI]6300/$\Ha = 0.22, meaning that
 F06296$+$5743 is an AGN-like galaxy, according to the criteria suggested by
 Armus, Heckman \& Miley ( 1989 ).

 Comparison with the power-law photoionization models of Ferland \& Netzer 
 ( 1983 ) shows that the models do not have line ratios simultaneously 
consistent with both the $\rm [NII]/$\Ha~ and the $\rm [OI]6300/$\Ha~ observed 
in F06296$+$5743. The shock models of Binette, Dopita, \& Tuohy ( 1985 ), 
however, can closely reproduce the relative emission line strengths seen in
 F06296$+$5743, implying that a starburst-induced, supernovae-driven superwind
( see, e.g. McCarthy, Heckamn \& van Breugel 1987 ) could contribute to the 
energy budget of this source rather than  a AGN.

\subsection{Nuclear Stellar Content}

   \begin{table}
      \caption{Line Intensities of F06296+5743}
      \begin{flushleft}
      \begin{tabular}{cclrr}
      \hline
      \hline
      $\lambda$ & $\lambda_{0}$ & Ion & EW & I/I(\Hb)\\
      (\AA) & (\AA) & & (\AA) &  \\
      \hline
      3965.9  &  3767.7  & H$_{11}$    &    9.9    & -1.269  \\
      3996.3  &  3796.5  & H$_{10}$    &    6.4    & -0.902   \\
      4036.6  &  3834.8  & H$_{9}$     &    7.4    & -0.946   \\
      4093.5  &  3888.9  & H${\zeta}$  &    9.6    & -1.376  \\
      4139.7  &  3932.8  & CaII K      &    4.0    & -0.562   \\
      4178.2  &  3969.3  & \HE$+$CaII H &    9.1    & -1.301  \\
      4317.9  &  4102.0  & H${\delta}$ &    8.3    & -1.022  \\
      4568.6  &  4340.2  & \Hr         &    8.2    & -1.064  \\
      5117.4  &  4861.6  & \Hb         &    8.2    & -1.00    \\
      5445.1  &  5172.9  & Mg I b      &    2.5    & -0.304   \\
      6200.8  &  5890.8  & Na I D      &    2.8    & -0.296   \\
      6629.9  &  6298.4  & [O I]       &   -2.8    & 0.265    \\
      6891.9  &  6547.4  & [N II]      &   -4.4    & 0.440    \\
      6907.7  &  6562.4  & \Ha         &  -12.4    & 1.227   \\
      6930.5  &  6584.0  & [N II]      &  -13.3    & 1.322   \\
      7067.7  &  6714.4  & [S II]      &   -3.4    & 0.290    \\
      7084.4  &  6730.2  & [S II]      &   -4.7    & 0.391    \\
      \hline
      \end{tabular}
      \end{flushleft}
      \end{table}

As is well known that knowledge of the stellar content is basic for
understanding star formation in galaxies. In the spectrum of F06296$+$5743, we
find several absorption lines which are discriminator for stellar population,
they are Balmer lines, and CaII K line ( see, e.g. O'Connell 1973,
Heckman 1980, Armus, Heckman \& Miley 1989 ).

\subsubsection{Balmer Lines}
In older stellar population dominated by G-K giants, the absorption at \Hb~ has
been measured as 1-2\AA ( Kennicutt 1983 ). The Balmer equivalent widths in A 
types, which are the largest ones in various type of stars, typically have 4\AA~
at \Hb~ as giants, and $\sim$ 8\AA~at \Hb~ as dwarfs ( O'Connell 1973 ). The
measured Balmer equivalent widths of F06296$+$5743 at \Hb, \Hr, and \Hd, see
column 4 of Table 1, are 8.2, 8.2 and 8.3\AA, respectively, indicating
dominant population of A dwarfs in the nuclear region of F06296$+$5743.

The width of Balmer lines provides additional evidence for the identification of
a young, and probable dwarf, component in the nucleus of this source. 
It is known that pressure broadening of the Balmer lines is greatest in dwarfs.
For F06296$+$5743, the \Hd's width at 20\% of the maximum negative intensity 
is about 28\AA, corrected for the instrumental profile. This value fits the
correlation between the equivalent width of \Hd~ and its width well ( see, e.g.
Heckman 1980 ), suggesting again that the Balmer lines come primarily from
young, and probable dwarfs, stars rather than a hotter giants or horizontal
branch.

\subsubsection{CaII K Line}

The K line of $\lambda$3933 is known to be a useful indicator for spectral types
of A and later. The criteria for HD spectral classification in A type stars
( Lang 1980 ) are as follows: CaII K=0.4 \Hd~ for A2 type; K=0.8 \Hd~ for A3;
K $>$ \Hd~ for A5 type. The measured intensity of CaII K line in F06296$+$5743,
see column 5 of Table 1,is about half of that of \Hd, suggesting the existence 
of a substantial content of A2-A3 stars in the nucleus.

Vacca \& Conti ( 1992 ) have estimated strength of interstellar CaII K feature,
which is not greater than 0.15\AA. Thus, the observed CaII K line in 
F06296$+$5743 is primarily of stellar origin.

Besides, the K line index ( O'Connell 1973 ) derived from the spectrum of this
source is about 0.25, a value relavant to A3-A5 type stars. This is consistent
with the inference from comparison of CaII K with \Hd~ that the dominant
population in the nuclear region might be A2-A3 stars.

\subsubsection{"V/R" and Continuum Colors}

French ( 1980 ) has introduced continuum colors,  B'-V' and V'-R', defined at
4200\AA(B'), 5300\AA(V') and 7000\AA(R') respectively, to set the dominant 
sources for continuum energy distribution of extragalactic objects. According
to his definition, we derived the continuum colors of F06296$+$5743, 
B'-V'=-0.20, V'-R'=-0.32. Compared with the colors for several main-sequence 
stars provided by French ( 1980 ), we found that the colors of F06296$+$5743 
is somewhat later than A0 ( B'-V'=-0.25, V'-R'=-0.33 ), and earlier than A5
( B'-V'=-0.15, V'-R'=-0.22 ).

The continuum index V/R has been used for characterizing energy distribution of
galaxies ( O'Connell 1973 ). It is also found of great use to set the dominant
contribution to the spectral energy budget of F06296$+$5743. The observed 
dereddened V/R of our source was obtained to be $-$0.32. The data for
F06296$+$5743 did not extend to 7400\AA~in the object's frame, and the continuum
had to be extrapolated about 200\AA~to derive R'. But the errors introduced by
this extrapolation may be small. The observed V/R will not be larger than -0.29 
in any case.

Due to the tight correlation between the V/R continuum index and the V$-$R 
broad-band color below V/R = +0.15 ( O'Connell 1973 ), we could apply the 
K-correction, $\rm K_{V}-\rm K_{R}$, by Whitford ( 1971 ) to V/R. Its value is
about 0.045 for z=0.05262. The final, dereddened V/R would be roughly in 
between -0.29 and -0.37, a value corresponding to B8-A0 energy distribution.
( O'Connell 1973 ).

In summary, the spectral analyses described in section 3.2.1 through section 
3.2.3 come to the more or less same conclusion that the dominant population of 
F06296$+$5743 might be early A type stars, A2/A3 most likely.

\subsubsection{Age of Starburst}

The derredened broad-band color V$-$R of F06296$+$5743 could be roughly obtained
from the tight correlation between the continuum index V/R and the broad-band
color V$-$R below V/R = +0.15 ( O'Connell 1973 ), as we pointed out in sec. 3.2.

The V$-$R derived this way is in between -0.03 and +0.11 at most, corresponding 
to V/R of -0.29 $-$ -0.37. By comparison with the predictions of V$-$R from
Leitherer \& Heckman ( 1995 ) for a continuous star-formation rate, the burst 
age would be $< 7\times 10^{7}$ yr  in any case.

The correction for the intrinsic reddening, if we could do that, would make
V$-$R bluer than the above values. So, it would be safe for us to estimate the
age of starburst to be $< 10^{8}$yr.

\section{Conclusion}

It is interesting to find that despite the very strong internal reddening of
F06296$+$5743, inferred from its high far-infrared luminosity of
log $\rm L_{FIR}=10.69$ with Hubble constant of 75$\rm km/sec/Mpc$, the dominant
nuclear stellar population of ealy type of A stars deduced from the continuum 
color, V/R or V$-$R, is basically matching what derived from the Balmer line 
intensities and line ratios. This would be a strong evidence to indicate that
the contribution from old population within the galactic bulge has been 
substantially supressed even at an age of burst $\sim$ 7$\times 10^{7}$ yr.

We have noticed that in various empirical models discussed by Bica, Alloin, \&
Schmidt ( 1990 ), the model value of the continuum color, V$-$R, is 0.38 for
10\% mass burst at age of 7$\times 10^{7} - 2\times 10^{8}$ yr. It is much
redder than the derredened color of 0.11 for F06296$+$5743. We have also 
noticed, however, that the predicted continuum colors, V$-$R, are getting bluer 
with increasing of the burst strength in models presented by  Bica, Alloin, \&
Schmidt ( 1990 ). It would be very reasonably and naively then to suspect that 
the situation showing in F06296$+$5743 might be a result from a very massive
starburst, more than 10\% burst strength at least.

\begin{acknowledgements} 
We would like to thank the anonymous referee for his valuable criticism and
suggestions. This work was supported by grants from National Science and
Technology Commission and National Natural Science Foundation.
\end{acknowledgements}


\begin{thebibliography}{}

\bibitem{}Armus L., Heckman T. M. \& Miley G. K. 1989, ApJ, 347,727.
\bibitem{}Barnes, J.V., \& Hayes, D.S. 1984, IRS Standard Star Manual
\bibitem{}Bica E., Alloin D., \& Schmidt A. 1990, MNRAS, 242, 241
\bibitem{}Binette, L., Dopita, M.A. \& Tuohy, I.R. 1985, ApJ, 297, 476
\bibitem{}Burstein, D., \& Heiles, C. 1984, ApJS, 54, 33
\bibitem{}Ferland, E.D., \& Netzer, H. 1983, ApJ, 264, 105
\bibitem{}French, H.B. 1980, ApJ, 240, 41
\bibitem{}Heckman, T.M. 1980, A\&A, 87, 142
\bibitem{}Kennicutt, R.C. 1983, ApJ, 272, 54
\bibitem{}Kurucz, R.L. 1979, ApJS, 40, 1
\bibitem{}Lang, K.R. 1980, Adtrophysical Formulae, Springer-Verlag
\bibitem{}Leitherer C. \& Heckman T. M. 1995, ApJS, 96, 9
\bibitem{}McCarthy, P.J., Heckman, T.M., \& van Breugel, W. 1987, AJ, 93, 264
\bibitem{}O'Connell, R.W. 1973, AJ, 78, 1074
\bibitem{}Osterbrock, D.E., Shaw, R.A. \& Veilleux, S. 1990, ApJ, 352, 561
\bibitem{}Vacca, W.D., \& Conti, P.S. 1992, ApJ, 401, 543
\bibitem{}Whitford, A.E. 1971, ApJ, 169, 215

\end{thebibliography}
\end{document}